\pdfoutput=1

\documentclass[pmlr]{jmlr}

\usepackage{longtable}
\usepackage[tableposition=top]{caption}

\usepackage{floatrow}
\usepackage{bbm}
\usepackage{amsmath,amssymb,amsfonts}
\usepackage{algorithmic}
\usepackage{graphicx}
\usepackage{textcomp}
\usepackage{xcolor}
\usepackage{multirow}
\usepackage{booktabs}
\usepackage{bbold}
\usepackage{mathtools}
\usepackage{amsfonts}
\usepackage{amsmath}
\usepackage{amssymb}
\usepackage{enumitem}
\usepackage[switch]{lineno}

\usepackage{booktabs}

\usepackage{setspace}
\usepackage{longtable}

\usepackage[load-configurations=version-1]{siunitx} 

\makeatletter
\def\set@curr@file#1{\def\@curr@file{#1}} 
\makeatother

\theorembodyfont{\upshape}
\theoremheaderfont{\scshape}
\theorempostheader{:}
\theoremsep{\newline}

\jmlrvolume{126}
\jmlryear{2021}
\jmlrworkshop{Machine Learning for Healthcare}

\usepackage{floatrow}
\newfloatcommand{capbtabbox}{table}[][\FBwidth]

\title[Mind the Performance Gap - Dataset Shift During Prospective Validation]{Mind the Performance Gap:\\ Examining Dataset Shift During Prospective Validation}

\author{\Name{Erkin \"{O}tle\d{s}$^{\,*\,1,2}$}
    \Email{\href{mailto:eotles@umich.edu}{\color{black}{eotles@umich.edu}}}
\AND
    \Name{Jeeheh Oh$^{\,*\,3}$}
    \Email{\href{mailto:jeeheh@umich.edu}{\color{black}{jeeheh@umich.edu}}}
\AND
    \Name{Benjamin Li$^{\,2}$}
\AND 
    \Name{Michelle Bochinski$^{\,4}$}
\AND
    \Name{Hyeon Joo$^{\,5,6}$}
\AND
    \Name{Justin Ortwine$^{\,5}$}
\AND
    \Name{Erica Shenoy$^{\,7}$}
\AND
    \Name{Laraine Washer$^{\,8}$}
\AND
    \Name{Vincent B. Young$^{\,8,9}$}
\AND
    \Name{Krishna Rao$^{\,8}$}
\AND
    \Name{Jenna Wiens$^{\,3}$}
    \Email{\href{mailto:wiensj@umich.edu}{\color{black}{wiensj@umich.edu}}}
\AND \addr {\begin{singlespace}\noindent\footnotesize $^*$\!Authors of equal contribution. \\ 
            $^1$\!Department of Industrial \& Operations Engineering, University of Michigan; 
            $^2$\!Medical Scientist Training Program, University of Michigan Medical School;
            $^3$\!Division of  Computer Science \& Engineering, University of Michigan;
            $^4$\!Nursing Informatics, Michigan Medicine;
            $^5$\!Department of Anesthesiology, University of Michigan Medical School;
            $^6$\!Department of Learning Health Sciences, University of Michigan Medical School;
            $^7$\!Infection Control Unit \& Division of Infectious Diseases, Massachusetts General Hospital \& Harvard Medical School;
            $^8$\!Department of Internal Medicine - Division of Infectious Diseases, University of Michigan Medical School;
            $^9$\!Department of Microbiology \& Immunology, University of Michigan Medical School
            \end{singlespace}
        }
}

\newcommand{\eg}{\textit{e.g.}, }
\newcommand{\ie}{\textit{i.e.}, }

\newcommand{\yrab}[2]{'{#1}-'{#2}}
\newcommand{\ci}{95\% CI: }

\begin{document}

\maketitle

\begin{abstract}
Once integrated into clinical care, patient risk stratification models may perform worse compared to their retrospective performance. To date, it is widely accepted that performance will degrade over time due to changes in care processes and patient populations. However, the extent to which this occurs is poorly understood, in part because few researchers report prospective validation performance. In this study, we compare the 2020-2021 (\yrab{20}{21}) prospective performance of a patient risk stratification model for predicting healthcare-associated infections to a 2019-2020 (\yrab{19}{20}) retrospective validation of the same model.  We define the difference in retrospective and prospective performance as the \textit{performance gap}. We estimate how i) ``temporal shift'', \ie changes in clinical workflows and patient populations, and ii) ``infrastructure shift'', \ie changes in access, extraction and transformation of data, both contribute to the performance gap. Applied prospectively to 26,864 hospital encounters during a twelve-month period from July 2020 to June 2021, the model achieved an area under the receiver operating characteristic curve (AUROC) of $0.767$ (95\% confidence interval (CI): $0.737$, $0.801$) and a Brier score of $0.189$ (\ci$0.186$, $0.191$). Prospective performance decreased slightly compared to \yrab{19}{20} retrospective performance, in which the model achieved an AUROC of $0.778$ (\ci$0.744$, $0.815$) and a Brier score of $0.163$ (\ci$0.161$, $0.165$). The resulting performance gap was primarily due to infrastructure shift and not temporal shift. So long as we continue to develop and validate models using data stored in large research data warehouses, we must consider differences in how and when data are accessed, measure how these differences may negatively affect prospective performance, and work to mitigate those differences.

\end{abstract}

\section{Introduction}

To date, the application of machine learning (ML) for patient risk stratification in clinical care has relied almost entirely on ``retrospective'' electronic health record (EHR) data \citep{RN723, RN881}. That is, researchers typically train and validate models using data sourced from a database \textit{downstream} from data used in clinical operations (\textit{e.g.}, a research data warehouse or MIMIC III  \citep{RN723}). These data are extracted, transformed and stored in an effort to serve researchers without interrupting hospital operations. While critical to initial model development, model evaluation using such data may not be representative of prospective model performance in clinical practice. Importantly, it is \textit{prospective} or ``real-time'' model performance that ultimately impacts clinical care and patient outcomes \citep{RN869, RN879}. Although retrospective performance serves as an approximation of prospective performance, the validity of such an approximation relies on the assumption that the two datasets come from the same distribution (\ie the datasets have no major differences in the relationships of covariates and outcome). However, many ML models are developed and validated with datasets that do not accurately represent  their intended prospective use \citep{RN882}. Without prospective evaluation, it is impossible to estimate \textit{a priori} how a model will perform when deployed. 

The need for prospective validation has been previously recognized in the context of screening for diabetic retinopathy \citep{RN860, RN861, RN686}. However, these studies rely mostly on imaging data and as a result the difference in infrastructure for model development and deployment is minimal. With respect to models that rely on structured EHR data, researchers have started to report prospective performance. For example, \citet{RN859} prospectively compared a model to predict in hospital resuscitation events with existing standards of care (\textit{e.g.}, rapid response team activation). In addition, \citet{RN432} prospectively validated an in-hospital mortality prediction model. While these studies make an important step towards model integration in clinical care, they do not specifically assess the root cause of discrepancies between prospective and retrospective performance.

To date, factors driving the differences between prospective and retrospective model performance have been largely attributed to changes in clinical workflow \citep{RN888, RN874} or patient populations \citep{RN889, RN1013}. For example, a global pandemic might lead to differences in personal protective equipment protocols. This change in gowning and gloving may have an impact on communicable diseases within the hospital, and this in turn may affect model performance. However, such changes are  difficult, if not impossible, to anticipate a year prior to outbreak \citep{RN995, RN996}.

Here, we compare the effects of ``temporal shift'' (\ie changes due to differences in clinical workflows and patient populations) on model performance to another kind of shift: ``infrastructure shift.'' We define infrastructure shift as changes due to differences in the data extraction and transformation pipelines between retrospective and real-time prospective analyses. For example, some data available retrospectively may not be available prospectively because of the processing pipeline at one's institution (\eg vitals might be backdated by the clinical care team). Differences in how the data are sourced and preprocessed between retrospective and prospective pipelines may be more systematically addressed if recognized. However, it is currently unknown to what extent degradation in prospective performance can be attributed to changes in temporal shift vs. infrastructure shift. 

In this paper, we explore the prospective validation of a data-driven EHR-based patient risk stratification tool for predicting hospital-associated \textit{Clostridioides} \textit{difficile} infections (CDI) at University of Michigan Health, a large tertiary care academic health system. CDI is associated with increased length of stay and hospital costs and considerable morbidity and mortality \citep{RN832, RN883, RN884, RN885}. The ability to accurately predict infections in advance could lead to more timely interventions, including patient isolation and antibiotic stewardship strategies, curbing the incidence and spread of infection. We measure the \textit{performance gap} between prospective and retrospective pipelines. More specifically, we \textit{quantify} how much of the performance gap can be attributed to temporal and infrastructure shift. 


\subsection*{Generalizable Insights about Machine Learning in the Context of Healthcare}
As the field of machine learning for healthcare advances and more models move from `bench' to `bedside,' prospective validation is critical. However, the majority of machine learning models are developed using retrospective data. We explore the impact this disconnect can have on the performance gap (\ie the difference between prospective performance and retrospective performance) through a case study in which we prospectively validated an EHR-based patient risk stratification model for CDI. Our contributions are as follows:

\begin{itemize}
    \item We formalize the notion of performance gap when validating ML-based models in clinical care.
    \item We characterize the differences between a retrospective pipeline and a prospective pipeline and the resulting impact on the performance gap. 
    \item We quantify how much of the performance gap can be attributed to temporal shift and infrastructure shift.
    \item We develop methods and approaches to identify contributors to the performance gap.
    \item We propose potential solutions for mitigating the effects of differences in retrospective versus prospective data infrastructure on the performance gap.
\end{itemize}

We do not present a new machine learning algorithm or architecture, but instead share insights gained through our experience with developing and validating an EHR-based model for patient risk stratification, highlighting practical considerations for those who are moving from the `retrospective' to `prospective' setting. Given that the ultimate goal of machine learning for healthcare is to improve patient care, early considerations regarding prospective validation are critical to ensuring success.

\section{Methods}

\newcommand{\X}{\mathbf{X}}
\newcommand{\y}{\mathbf{y}}

\newcommand{\D}{\mathcal{D}}
\newcommand{\Dret}{\D_{ret}}
\newcommand{\Dpro}{\D_{pro}}

\begin{figure}[h]
  \centering 
  \includegraphics[width=4in]{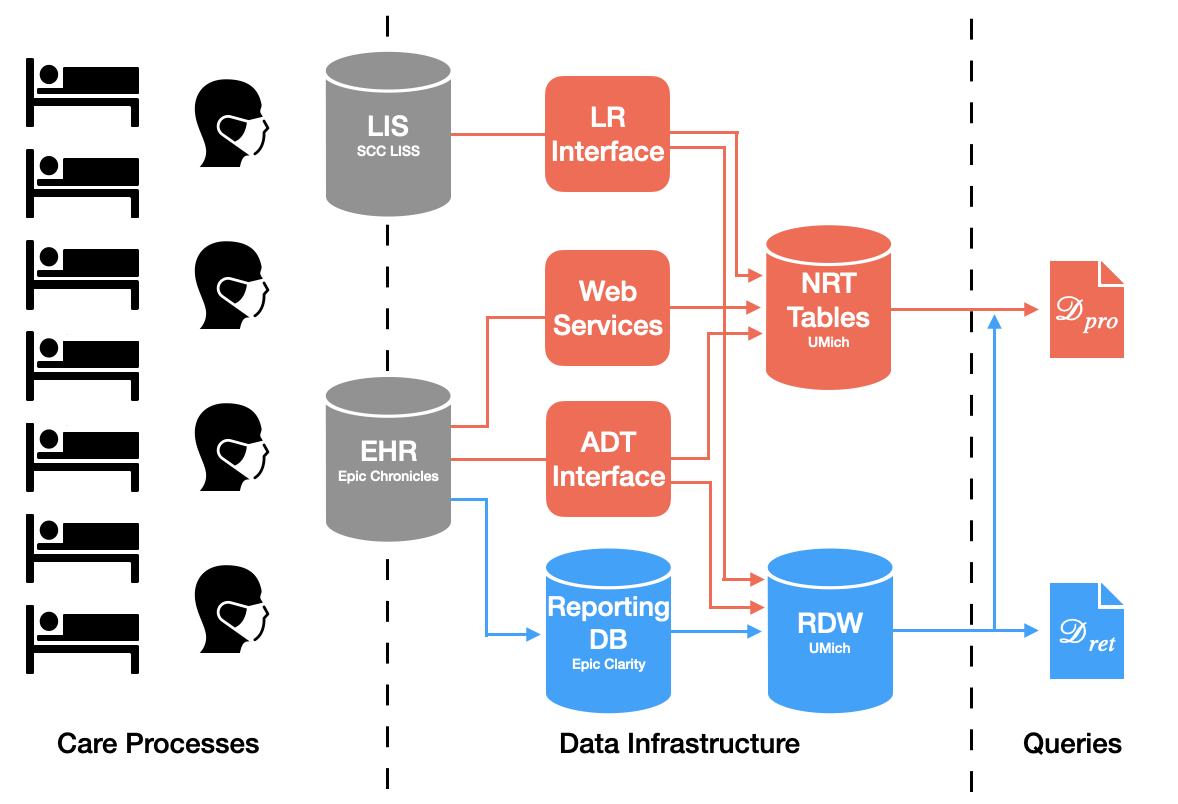} 
  \caption{Prospective and Retrospective Pipelines.  Information generated from the care process is documented in the electronic health record (EHR), produced by Epic Systems Corporation, and laboratory information system (LIS), produced by Soft Computer Consultants (SCC). Data are extracted from these sources using two different pipelines. For EHR data the near real-time prospective pipeline is primarily based on a web service architecture and has its information stored in near real-time (NRT) database tables. It extracts data from the EHR more frequently, with less lead time and processing, allowing for prospective implementation of predictive models (\ie it produces prospective datasets, $D_{pro}$). The bottom pipeline is a retrospective data pipeline that extracts data less frequently, but with more curation and processing (\ie it generates large retrospective datasets, $D_{ret}$).
  Both pipelines rely on an lab results (LR) interface that passes information from the LIS and an admission, discharge, and transfer (ADT) interface that passes admission information from the EHR. Components in the pipeline that can be interacted with in near real-time (\ie prospectively) are depicted in red. Components in which subsets of data require time to pass before having complete information (\ie retrospectively) are depicted in blue.
  The near real-time query utilizes historical patient information, although this information is technically collected via the retrospective pipeline, it is considered up-to-date when queried by the near real-time query.
  }
  \label{fig:pipelines} 
\end{figure}

\textbf{Overview.} In the context of inpatient risk stratification, we present an evaluation framework to quantify the differences in performance between machine learning models applied in real-time and the anticipated performance based on retrospective datasets. In framing this problem, we examine two major sources of differences: 1) the shift in the relationships between the features and labels over time due to changes in clinical workflows and patient populations (\ie temporal shift) and, 2) the difference in the infrastructure for extracting data retrospectively versus prospectively (\ie infrastructure shift). We leveraged a previously developed and retrospectively validated framework for predicting hospital-associated CDI in adult inpatients \citep{RN312, RN311}. Below, we briefly describe how this framework was applied retrospectively to develop and validate a model, and then updated to apply prospectively such that model predictions were generated for all adult inpatients on a daily basis. We also share important details regarding the data extraction and processing pipelines available for model development and prospective implementation at University of Michigan Health (summarized in \textbf{Figure \ref{fig:pipelines}}). Though some aspects (\eg the precise downstream research database) may be unique to our institution, many aspects of the data pipelines are generalizable.\\

\subsection{Study Cohort}
This retrospective and prospective cohort study was approved by the University of Michigan Institutional Review Board. Our study population included all adult hospital patient encounters (\ie inpatient admissions) from January 2013 through June 2021 to University of Michigan Health. University of Michigan Health has over 1,000 beds and is the tertiary care academic health center associated with the University of Michigan. Because we were interested in primary, non-recurrent, hospital-associated CDI we excluded encounters with a length of stay less than three calendar days and individuals who tested positive in the first two calendar days of the encounter or in the proceeding 14 days prior to the hospital encounter \citep{RN1023}.  

\subsection{Prediction Task}
The task was formulated as a binary classification task where a patient encounter was labeled 1 if the patient tested positive for CDI during the encounter and 0 otherwise. The diagnosis of CDI was identified using a tiered approach, reflecting the institution’s \textit{C. difficile} testing protocol when clinicians obtained stool samples for \textit{C. difficile} based on clinical suspicion for active disease. First, samples were tested using a combined glutamate dehydrogenase antigen enzyme immunoassay and toxin A/B EIA (C. Diff Quik Chek Complete, Alere, Kansas City, MO). No further testing was needed if results were concordant. If discordant, a secondary polymerase chain reaction (PCR) for the presence of toxin B gene (GeneOhm Cdiff Assay, BD, Franklin Lakes, NJ) was used to determine the outcome. That is, if positive by PCR the encounter was considered a CDI positive case. We make predictions daily, with the goal of identifying high-risk patients as early as possible during an encounter and prior to their diagnosis.

\subsection{Model Development}

\noindent
\textbf{Training Data.} Our training cohort included patients admissions between 2013-2017 who met our inclusion criteria. When applying our inclusion criteria, we relied on patient class codes to identify hospitalized patient encounters. For each patient admission included in the training data, we extracted a binary classification label and information pertaining to a patient's demographics, medical history, laboratory results, locations, vitals, and medications. Once retrospective data were extracted, we transformed the data into $d$-dimensional binary feature vectors representing each day of a patient's admission (\ie an encounter-day). Features were transformed into a binary representation. Categorical features were transformed using one-hot encoding. Real-valued (numerical) features were split into quintiles and then also one-hot encoded. This is described in further detail in the feature preprocessing section of \citet{RN311} and in \textbf{Supplemental Material A}. \\

\noindent
\textbf{Training Details.}
We employ a modeling approach previously described and validated at multiple institutions \citep{RN311,RN312}. In brief, we used a logistic regression model that uses a multitask transformation of the inputs in order to learn time varying parameters \citep{RN991}. The multitask regularized logistic regression model seeks to learn an encounter level label (\ie if the patient is ever diagnosed over their entire encounter). It does so by minimizing the cross-entropy loss at the encounter-day level.
We subsampled encounter-days to reduce bias towards patient encounters with longer lengths of stay. This was done by randomly selecting 3 encounter-days per encounter (our inclusion criteria dictates that all encounters will have a length of stay of at least 3 days). This ensured that all encounters were represented by an equivalent number of encounter-days. Cross validation folds were determined by year to partially account for dataset shift. Hyperparameters were selected based on cross-validation across years in the training data optimizing for the AUROC. This approach is described in detail in \citet{RN311} and \citet{RN312}.

\subsection{Data Infrastructure}

\noindent
\textbf{Retrospective Pipeline.} Data used for model development and retrospective validation were extracted from a research data warehouse (RDW) at the University of Michigan. These data were extracted, transformed, and loaded (ETL) from University of Michigan Health's Epic EHR instance (Epic Systems Corporation, Verona, WI) and laboratory information system (LIS, Soft Computer Consultants, Clearwater, FL). More precisely, the majority of EHR data were extracted nightly from the EHR's underlying Epic MUMPS-based Chronicles database and then transformed and loaded into our instance of Epic Clarity, a SQL database. A second ETL process was then carried out, with data passed to a second SQL database, RDW. RDW is based on a health information exchange data model (initially developed by CareEvolution, Ann Arbor, MI), however, in order to support research operations, has undergone significant additional custom design, development, and maintenance by the RDW development team at the University of Michigan. The timing of this second ETL process varied. However, the total delay between data being entered in the EHR to it arriving in RDW typically ranged between 1 day and 1 week. In addition to this data pipeline, our EHR also passes hospital occupancy information directly to RDW via an admission, discharge, and transfer (ADT) interface. Finally, RDW also captures information from the LIS using a lab results (LR) interface. RDW is designed for large queries of transformed EHR data. Thus, we refer to these data as `retrospective' since they were primarily used for retrospective analyses. 

\noindent
\\\textbf{Prospective Pipeline.} Not all data included in the retrospective model were available in near real-time through the pipeline described above (\eg medications or laboratory values for current encounters). Thus, we developed a near real-time prospective pipeline, which built upon the existing retrospective pipeline by adding daily updates (ETLs) of data that were previously unavailable in real-time. We developed custom EHR web services to update the data necessary for model predictions. Specialized near real-time (NRT) database tables were created to access medications, vital sign measurements, hospital locations, in near real-time; \ie with a delay of less than an 8-hours. This maximum delay corresponds with the maximum duration for the recording of non-urgent information into the EHR (\ie the length of a typical nursing shift). In conjunction with the EHR web services, laboratory results and admission information are passed to the NRT tables using the aforementioned LR and ADT interfaces respectively. Additionally, we continued to use components of the retrospective pipeline to extract features pertaining to historical patient data (\eg medications associated with previous encounters).

Overall, daily data extracts were inherently different from historical data and required careful validation to ensure queries were accessing the correct aspects of the EHR.  Once extracted, we applied an identical preprocessing step as described in the retrospective pipeline. Using these daily data streams, we generated daily risk scores for all adult hospital encounters in our study cohort. Model results were generated daily and stored on a secure server. These scores were not made available to any clinical care team members and were only accessed by the authors.

\subsection{Model Validation}
The model takes as input data describing each day of a patient's hospital encounter, extracted through either the retrospective or near-real time prospective pipeline describe above, (\textit{e.g.}, laboratory results, medications, procedures, vital sign measurements, and patient demographics) and maps these data to an estimate of the patient's daily risk of CDI. This estimate is updated over the course of the patient's hospital encounter with the goal of helping clinicians identify patients at risk of developing CDI over the remainder of the patient's hospital encounter. 
 
\noindent
\\\textbf{Retrospective Validation.}
We validated the model on data pertaining to patient hospital encounters from the study cohort from 2018-2020. We extracted these data using the retrospective pipeline and identified hospitalized patients using patient class codes as we did in the training data (see above for inclusion criteria). In our main analyses, we focus on performance during the more recent year \ie \yrab{19}{20}. For completeness, results from \yrab{18}{19} are provided in the \textbf{Supplemental Material D}. 

We measured the area under the receiver operating characteristics curve (AUROC) and the sensitivity, specificity, and positive predictive value when selecting a decision threshold based on the 95$^{th}$ percentile from \yrab{18}{19}. In addition we computed the Brier score, by comparing the max probability of the outcome (\ie risk score) during a patient's visit with their actual outcome. We calculated empirical 95\% confidence intervals on each test set using 1,000 bootstrap samples.

\noindent
\\\textbf{Prospective Validation.}
We applied the model prospectively to all hospital encounters from July 10th, 2020 to June 30th, 2021, estimating daily risk of CDI for all patients who met our inclusion criteria. We relied on the hospital census tables instead of the patient class codes to identify our study population in real-time. The hospital census table tracked all hospitalized patients in real-time and enabled reliable identification of patients who were in the hospital for three calendar days or more. 

We compared retrospective performance in \yrab{19}{20} to prospective performance in \yrab{20}{21}. We evaluated model performance in terms of discrimination and calibration using the same metrics described above. In addition, to account for seasonal fluctuations in CDI rates \citep{RN1012}, we further compared AUROC performance on a month-by-month basis. We compared monthly retrospective performance in \yrab{19}{20} to monthly prospective performance in \yrab{20}{21}. Although encounters may span across multiple months, encounters were grouped into month-years based on the date of admission.
Finally, given the large shift to care processes resulting from the onset of the COVID-19 pandemic, we conducted a separate follow-up analysis in which we compared model performance prior to and following March 2020 (\textbf{Supplemental Material C}).

\begin{figure}[h]
  \centering 
  \includegraphics[width=3.5in]{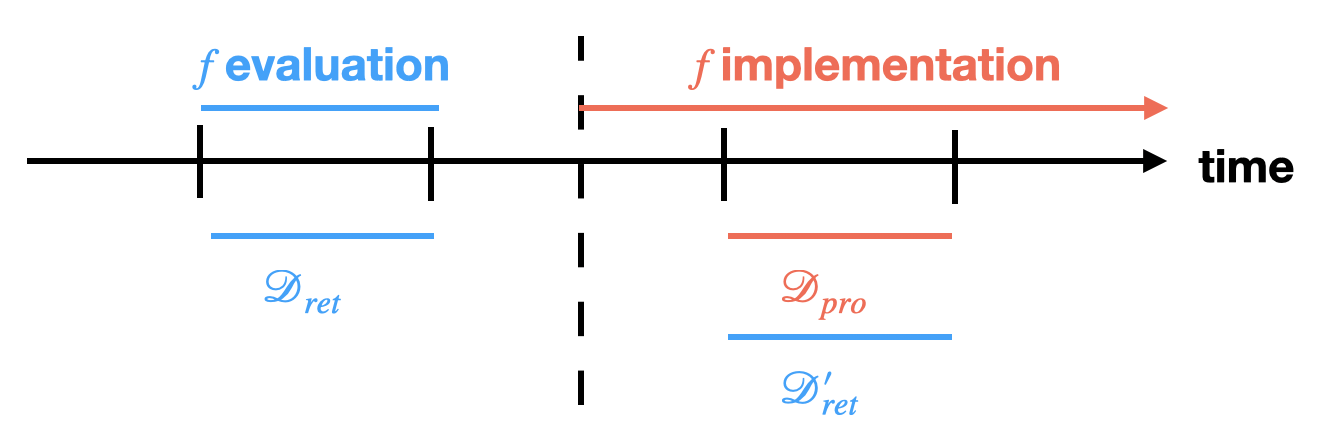} 
  \caption{Retrospective Evaluation and Prospective Implementation Timeline. The dashed vertical line denotes the time of model integration into silent prospective deployment. Prior to implementation, the retrospective pipeline may be used to retrospectively validate the model, $f$, applied to $\Dret$. After the model implementation the model is applied to $\Dpro$ using the prospective pipeline. Once sufficient time has elapsed the retrospective pipeline may be used to extract $\Dret'$, data from the same period of time as the prospective dataset.}
  \label{fig:timeline} 
\end{figure}

\subsection{Estimating the Performance Gap}

We evaluated model performance over time, comparing the same model applied to i) retrospective data from 2019-2020 and ii) prospective data from 2020-2021. We formalize our prospective validation framework as follows. 

Let the labeled data extracted from the retrospective and prospective pipeline be denoted as $\D_{ret}$ and $\D_{pro}$, respectively. Given a predictive model ($f$), $f$ maps a given feature vector ($\mathbf{X}$) to an estimate of patient risk ($y$), formally $f: \mathbf{X} \to y$. We evaluate $f$ applied to a dataset $\D$ using a performance measure function: $p: (f, \D) \to \mathbb{R}$ in which the goal is to maximize $p$. While $p(f, \D_{ret})$ often serves as an estimate for how the model will likely perform in the future, given differences between retrospective and prospective pipelines (discussed above) we do not necessarily anticipate $p(f, \D_{ret})$ to equal $p(f, \D_{pro})$. The difference between the retrospective and prospective model performance with respect to $p$ is the \textit{performance gap}:

\begin{equation}
\label{eq:1}
\Delta_p = p(f, \D_{ret})- p(f, \D_{pro}). 
\end{equation}

We analyzed the performance gap that arises between retrospective \yrab{19}{20} ($\D_{ret}$) and prospective \yrab{20}{21} ($\D_{pro}$). We measured the gap in terms of the AUROC because optimizing for discriminative performance was the primary goal of prior work. However, calibration is known to be  sensitive to temporal changes \citep{RN841, RN1001, RN874}. As such, we also measure the performance gap in terms of Brier Score \citep{RN302, RN989}. However, since one aims to minimize Brier Score and the performance gap assumes the goal is to maximize the performance measure, we take the negative of the Brier Score when computing the gap. Confidence intervals for the performance gap values were calculated using an empirical bootstrap where the samples ($1,000$ replications) were independently drawn for each data distribution separately.

When comparing model performance on retrospective data ($\D_{ret}$: \yrab{19}{20}) to model performance on prospective data ($\D_{pro}$: \yrab{20}{21}), there are two unique sources of potential differences: the shift in time period from \yrab{19}{20} to \yrab{20}{21} (\ie temporal shift) and the shift in pipelines (\ie infrastructure shift).

\begin{enumerate}

\item\textbf{Temporal Shift} arises due to changes in the underlying data generating process \textit{over time} (\eg the evolution of disease pathology, updates in clinical practice/workflows and changes in patient characteristics). If the retrospective and prospective pipelines were identical, then any difference between retrospective and prospective pipeline would be attributed to $\Delta_p^{time}$, defined as:
\begin{equation}
\label{eq:2}
\Delta_p^{time} = p(f, \D_{ret}) - p(f, \D_{ret}').
\end{equation} 
where we control for changes in infrastructure by re-extracting the data from the prospective period using the retrospective pipeline, $\D_{ret}'$.

\item\textbf{Infrastructure Shift} occurs when the data returned from the retrospective and prospective pipelines differ, after controlling for changes in time period. We calculate $\Delta_p^{infra}$ by comparing performance on the prospective dataset, $\D_{pro}$, to performance on a retrospective dataset generated for the identical time period $\D_{ret}'$ (\textbf{Figure \ref{fig:timeline}}). Once aligned in time, the only differences come about from the pipeline infrastructure used to create the datasets:
\begin{equation}
\label{eq:3}
\Delta_p^{infra} = p(f, \D_{ret}') - p(f, \D_{pro}).
\end{equation} 

\end{enumerate}

\noindent 
The performance gap from \textbf{Equation \ref{eq:1}} can be broken into these two sources of differences: 
\begin{equation}
\label{eq:4}
\notag
\begin{split}
\Delta_p &= p(f, \D_{ret})- p(f, \D_{pro}) \\
    &= p(f, \D_{ret})- p(f, \D_{pro}) + \Big( - p(f, \D_{ret}') + p(f, \D_{ret}' ) \Big) \\
    &= p(f, \D_{ret}) - p(f, \D_{ret}')   + p(f, \D_{ret}') - p(f, \D_{pro}) \\
    &=  \Delta_p^{time} + \Delta_p^{infra}.
\end{split}
\end{equation}

\noindent
To date, it is unknown to what extent differences in infrastructure contribute to the performance gap relative to differences that arise due to temporal shift. Thus, to control for temporal shift and estimate the effect of infrastructure shift on the performance gap, we used the retrospective pipeline to query data for the prospective validation time period (\ie July 10th, 2020 to June 30th, 2021), generating $\D_{ret}'$. Using the datasets $\Dret$, $\Dret'$, and $\Dpro$, we estimated $\Delta_p^{infra}$ and $\Delta_p^{time}$.

\subsection{Performance Gap Source Analyses}
To better understand how infrastructure choices related to the prospective pipeline contribute to $\Delta_p^{infra}$, we further analyzed $\Dpro$ and $\Dret'$. These analyses are motivated and described in the sections below.

\subsubsection{Analyzing Sources of Infrastructure Shift}
First, we characterized the difference in the estimated risk between the two datasets. Since the AUROCs and Brier Scores are summary statistics they may hide differences. Thus, we compared the risk scores output by the retrospective versus prospective pipeline for every encounter observed prospectively. These correspond to $\D_{ret}'$ and $\D_{pro}$, which share encounter-days, and thus the model's output for both datasets can be compared directly. Score pairs representing the maximum score were found for each encounter using the prospective pipeline and the retrospective pipeline. These score pairs were graphed as a scatter plot and then were analyzed for their concordance in terms of Pearson's correlation coefficient and the slope of the best-fit line. Extremely discordant prospective and retrospective score pairs were identified by selecting points far away from the best fit line (\ie score pairs with a difference $\geq 0.5$).

In order to understand factors that could potentially be addressed with modifications to infrastructure, we compared the pair of feature vectors present for each instance (a patient hospitalization encounter-day) in $\Dpro$ and $\Dret'$ by computing differences in feature inputs between the two datasets. The difference in the two data pipelines ($\Dpro$ and $\Dret'$) was quantitatively assessed for every feature, at the encounter-day level. Since our model utilized a binary feature space, we deemed features discrepant at the encounter-day level if their prospective and retrospective values were not exactly equivalent. This can be extended to real-valued (numerical) features through either exact equivalency or by using ranges. To assess the impact of these features, we stratified features by the absolute value of their model coefficients and the proportion of discrepancies.

Finally, large differences in features can result in minimal differences in estimated risk if the features that vary greatly are not deemed important by the model. Thus, we also calculated the effect of swapping out individual aspects of the prospective pipeline with the retrospective pipeline on overall prospective performance. These swaps were conducted on the feature groups  defined in \textbf{Supplemental Material A}. For every feature group we computed the performance of the model on a modified version of $\Dpro$ where the feature matrix $\mathbf{X}_{pro}$ has a column (or columns) corresponding to the feature replaced with values from $\mathbf{X}_{ret}'$. We conducted a feature swap analysis between the retrospective and prospective \yrab{20}{21} datasets using AUROC as the performance measure. Due to computational complexity, this analysis was only conducted at the mid-point of our study, and as such only uses data from July 10th to December 21st for both $\Dpro$ and $\Dret'$. 

\subsubsection{Analyzing Sources of Temporal Shift}
To determine sources of model performance degradation due to population and workflow changes over time we conducted a set of experiments that sought to uncover the impact of temporal shift by controlling for infrastructure shift sources. We identified sources of temporal shift by comparing the distribution of features between $\Dret$ and $\Dret'$. Specifically, for each feature we conducted a Z-test with a Bonferroni correction to test the difference in proportion of times that feature was 'turned on' in one time period versus the other, controlling for differences in infrastructure. \textit{E.g.}, was a particular medication used more frequently in one time period? We report the number of significant differences within each feature group (see \textbf{Supplemental Material A} for feature grouping).


\begin{table}[h]
    \centering
    \caption{Yearly Cohort Characteristics. Retrospective and prospective cohorts from \yrab{19}{20}, and \yrab{20}{21} each span from July 10th to June 30th of the following year. The cohorts have similar characteristics across years. For median values we also present the interquartile range (IQR).}
        \label{table:pop_characteristics}
    \begin{tabular}{@{}lll@{}}
    \toprule
                    &\yrab{19}{20}        & \yrab{20}{21}\\ 
                    &($\Dret$)            & ($\Dpro$)\\ 
                   &n=25,341        & n=26,864 \\
    \midrule
    Median Age (IQR)                        & 59 (41, 70)   & 60 (42, 71) \\
    Female (\%)                             & 51\%          & 51\%        \\
    Median Length of Stay (IQR)                        & 5 (4, 9)      & 5 (4, 9)    \\
    History of CDI in the past year (\%)    & 1.5\%         & 1.4\%       \\
    Incidence Rate of CDI (\%)              & 0.6\%         &  0.7\%       \\
    \bottomrule
    \end{tabular}
\end{table}

\begin{figure}[h]
\centering
    \subfigure[ROC Curves with 95\% Confidence Intervals. 
    \label{fig:validation_rocs}]{
        \includegraphics[width=2.75in]{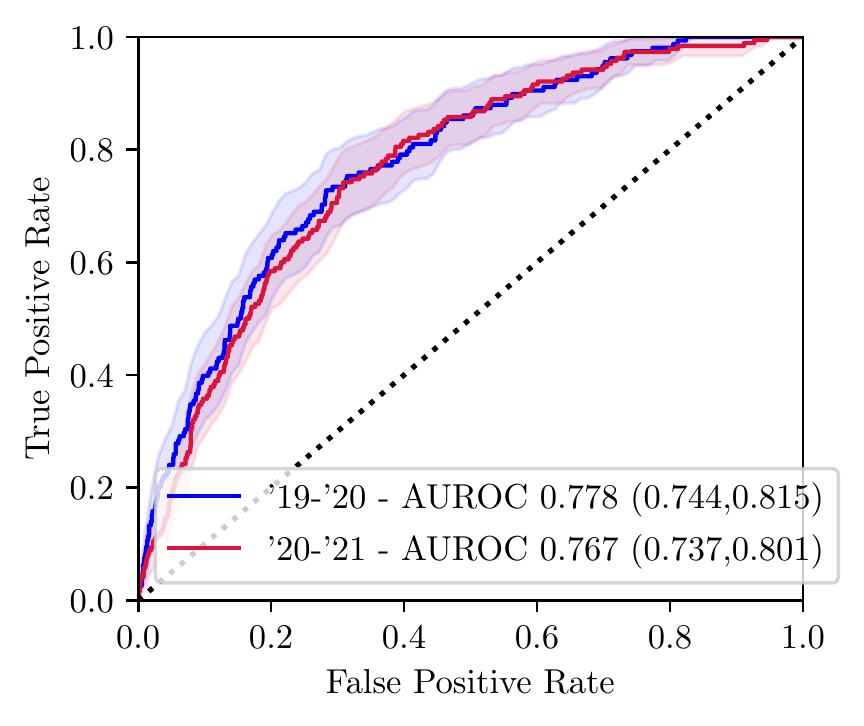}}
    \hspace{1em}{}
    \subfigure[Confusion Matrices and Performance Measures.  
    \label{fig:validation_confusion_matrices}]{
        \includegraphics[width=2.75in]{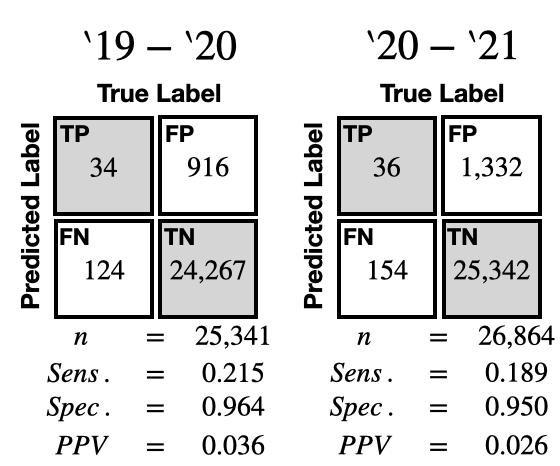}
    }
    \caption{Risk prediction model performance on \yrab{19}{20}, and \yrab{20}{21} validation datasets. When compared with the model's retrospective validation period (\yrab{19}{20}) performance the model demonstrated slightly worse discriminative performance, sensitivity, specificity, and positive predictive value during its prospective validation period (\yrab{20}{21}).}
    \label{fig:validation_performance}
\end{figure}

\begin{figure}[h!]
  \centering 
  \includegraphics[width=4.0in]{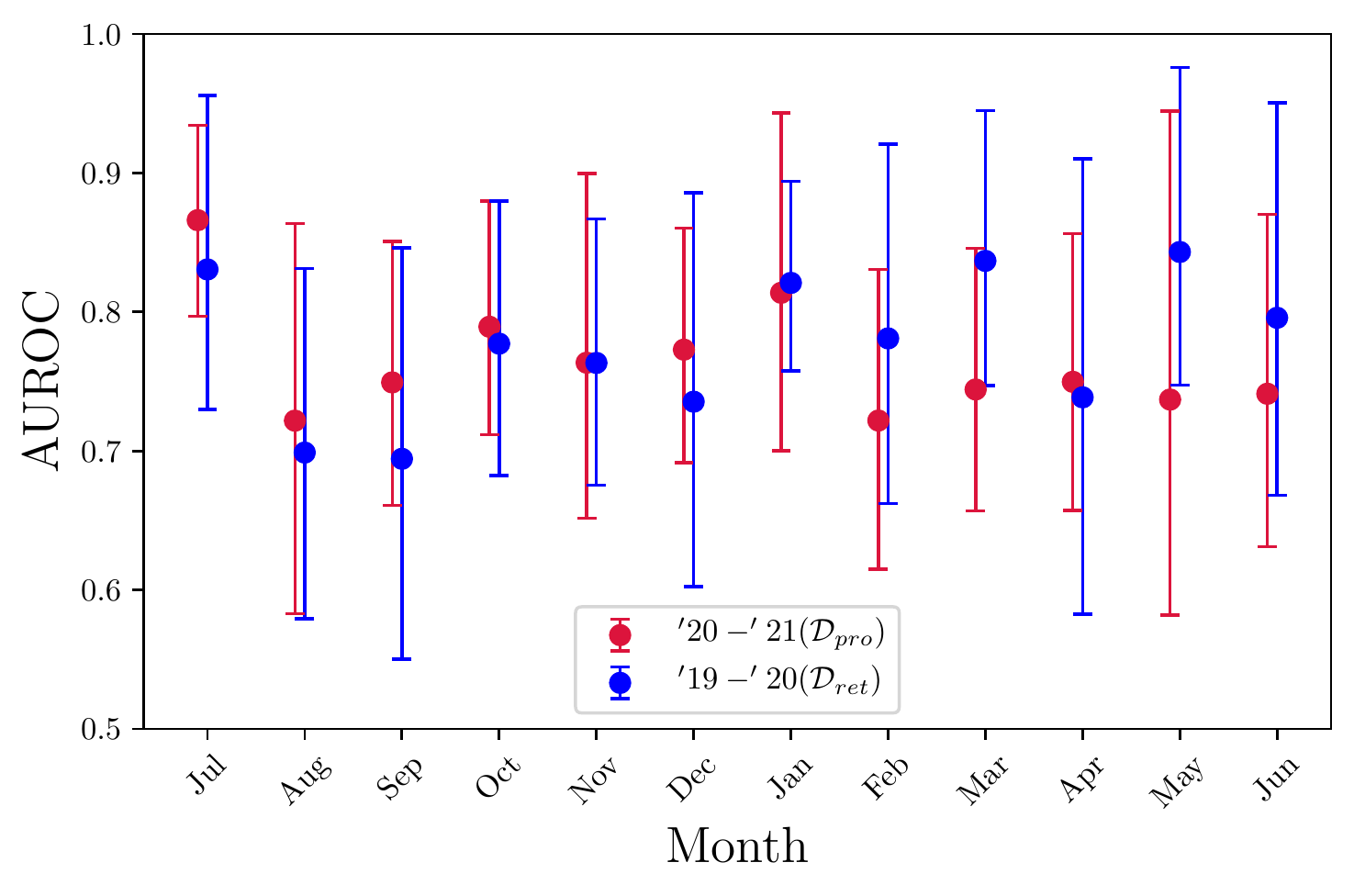} 
  \caption{Monthly AUROC Performance. AUROC for \yrab{20}{21} prospective dataset and the \yrab{19}{20} retrospective dataset broken down by month and bootstrap sampled $1,000$ times to generate empirical 95\% confidence intervals. Performance fluctuates month-by-month with the prospective pipeline generally outperforming or on par with retrospective performance with the exceptions of March and May.}
  \label{fig:monthlyauroc} 
\end{figure}

\newpage
\section{Results} 

Our training cohort included 175,934 hospital encounters, in which 1,589 (0.9\%) developed hospital-associated CDI. Feature extraction and processing resulted in 8,070 binary features. Our \yrab{19}{20} retrospective validation set ($\Dret$) consisted of 25,341 hospital encounters, in which 157 (0.6\%) met the CDI outcome. Prospectively in \yrab{20}{21}, we identified 26,864 hospital encounters, in which  188 (0.7\%) met the CDI outcome ($\Dpro$). Study population characteristics for both validation cohorts are reported in \textbf{Table \ref{table:pop_characteristics}}. During the prospective validation of the model, the prospective pipeline failed to run due to circumstances beyond the study team's control in 10 out of the 356 days. Specifically, from mid-December to February of 2021, an ADT data-feed issue led to a lag in some of the prospective data being processed. Risk scores were not generated on days in which the model failed to run. \\

\noindent
\textbf{Validation Results.} Applied to the \yrab{19}{20} and \yrab{20}{21} validation cohorts, the model achieved an AUROC of $0.778$ (\ci$0.744$, $0.815$) and $0.767$ (\ci$0.737$, $0.801$) and a positive predictive value of $0.036$ and $0.026$, respectively (\textbf{Figure \ref{fig:validation_performance}}). Model calibration was fair across both \yrab{19}{20} and \yrab{20}{21} datasets, Brier scores: 0.163 (\ci$0.161, 0.165$) and 0.189 (\ci$0.186$, $0.191$), respectively. On a monthly basis, prospective performance during \yrab{20}{21} did not differ significantly from the retrospective performance during \yrab{19}{20}, except in March and May (\textbf{Figure \ref{fig:monthlyauroc}}). \\


\begin{table}[h]
    \caption{Model Performance Comparison. The prospective validation ran from July 10th, 2020 to June 30th, 2021 (\yrab{20}{21}) and yielded dataset, $\Dpro$, and performance results. The \yrab{20}{21} retrospective dataset, $\Dret'$, uses the retrospective pipeline to pull the same population observed in $\Dpro$. The retrospective \yrab{19}{20} retrospective dataset pulled data from July 10th, 2019 to June 30th, 2020 in order to have an equivalent annual comparison. We see a positive AUROC performance gap and a negative Brier Score performance gap indicating degraded prospective performance.}\scalebox{0.85}{
    \begin{tabular}{@{}llll@{}}
    \toprule
                & \yrab{19}{20} Retrospective & \multicolumn{1}{l|}{\yrab{20}{21} Retrospective}      & \yrab{20}{21} Prospective \\
                & ($\Dret$)          & \multicolumn{1}{l|}{($\Dret'$)}              & ($\Dpro$) \\
                & n=25,341           & \multicolumn{1}{l|}{n=26,864}                & n=26,864 \\
    \midrule
    AUROC (\ci)      & 0.778 (0.744, 0.815) & \multicolumn{1}{l|}{0.783 (0.755, 0.815)}   &  0.767 (0.737, 0.801)     \\
    Brier Score (\ci) & 0.163 (0.161, 0.165) & \multicolumn{1}{l|}{0.186 (0.184, 0.188)} & 0.189 (0.186, 0.191) \\
    \bottomrule
    \end{tabular}
    }
    \centering
    \label{table:performance_gap}
\end{table}

\noindent
\textbf{Performance Gap.} Overall, the performance gap between $\mathcal{D}_{ret}$ in \yrab{19}{20} and $\mathcal{D}_{pro}$ in \yrab{20}{21} was $\Delta_{AUROC}=0.011$ (\ci$-0.033$, $0.056$)  and $\Delta_{Brier}=0.025$\footnote{Clarification: Reader may wonder why this isn't $0.026$, this is simply due to rounding.} (\ci$0.016$, $0.110$). Applied to the re-extracted retrospective \yrab{20}{21} cohort ($\mathcal{D}'_{ret}$) the model achieved higher discriminative and calibration performance, AUROC=$0.783$ (\ci$0.755, 0.815$) and Brier score=$0.186$ (\ci$0.184$, $0.188$). Thus, according to \textbf{Equations \ref{eq:1}, \ref{eq:2}, and \ref{eq:3}}, the performance gap breaks down as follows:


$\begin{matrix}
 & & & & & 
 \\
\Delta_{AUROC} & = &  
\text{AUROC}({\Dret}) - \text{AUROC}({\Dpro}) & = & 
\hphantom{-} 0.011 & 
(\text{\ci}-0.033, \; 0.056)
\\
\Delta_{AUROC}^{infra} & = &  
\text{AUROC}({\Dret'}) - \text{AUROC}(\Dpro) & =& 
\hphantom{-} 0.016 & 
(\text{\ci}-0.022, \; 0.058)
\\
\Delta_{AUROC}^{time} & = &  
\Delta_{AUROC} -\Delta_{AUROC}^{infra} & = & 
-0.005 & 
(\text{\ci}-0.051, \; 0.036)
\\
\\
\Delta_{Brier} & = & 
-(\text{Brier}(\Dret) - \text{Brier}(\Dpro)) & = & 
\hphantom{-} 0.025 & 
(\text{\ci}\hphantom{-}0.016, \; 0.110)
\\
\Delta_{Brier}^{infra} & = & 
-(\text{Brier}(\Dret') - \text{Brier}(\Dpro)) & = & 
\hphantom{-} 0.002 & 
(\text{\ci}-0.021, \; 0.064)
\\
\Delta_{Brier}^{time} & = &  
\Delta_{Brier} -\Delta_{Brier}^{infra}  & = & 
\hphantom{-} 0.023 & 
(\text{\ci}-0.003, \; 0.084)
\\
 & & & & &
\end{matrix}$

\begin{figure}[h]
  \centering 
  \includegraphics[width=4.0in]{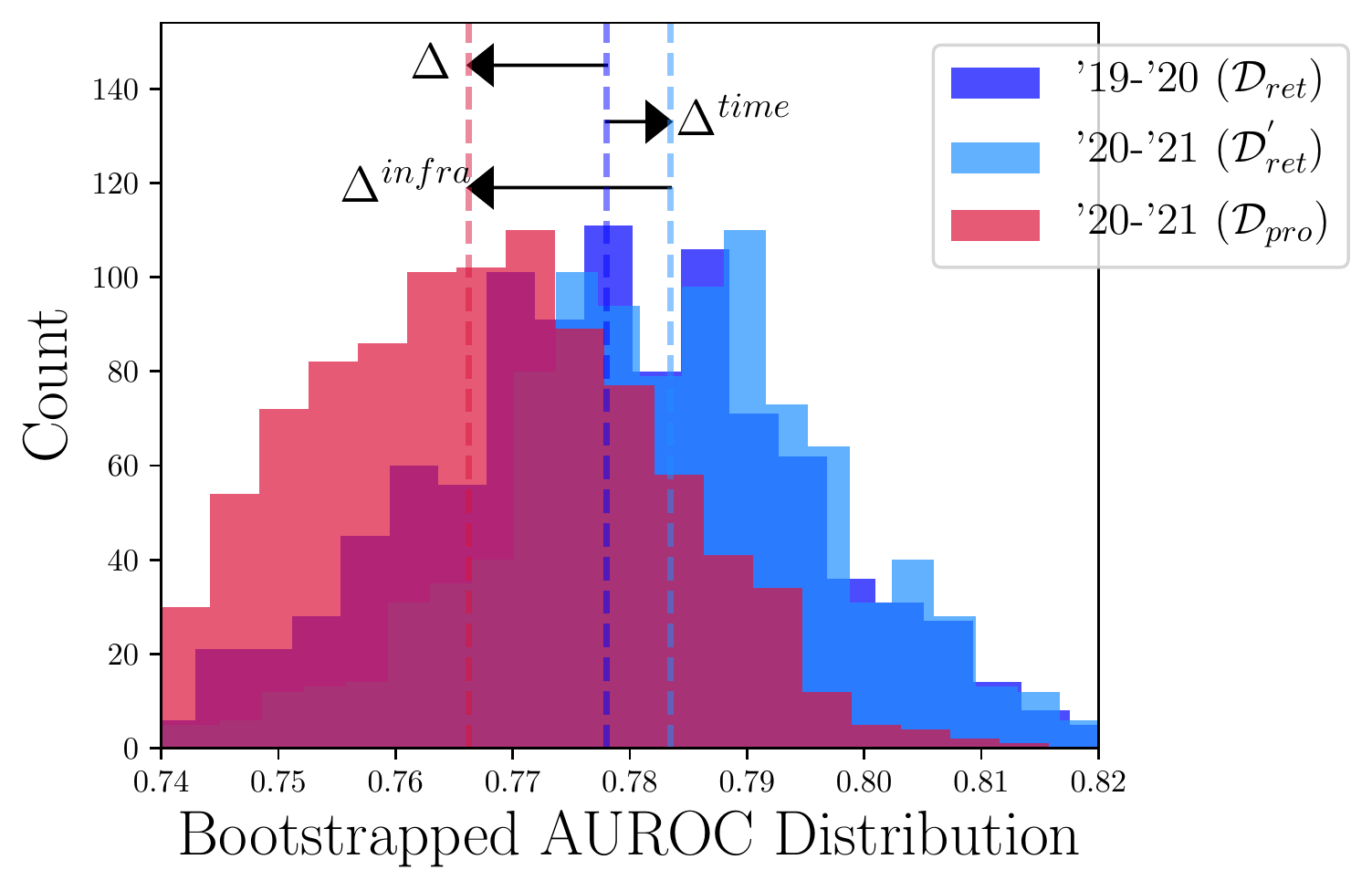} 
  \caption{Relationship between prospective risk scores and retrospective risk scoring. The bootstrapped distribution of the AUROCs of the model applied to three different datasets are shown along with the performance gap, $\Delta_{\text{AUROC}}$, and its components, $\Delta_{\text{AUROC}}^{infra}$, and $\Delta_{\text{AUROC}}^{time}$. The overall gap is positive demonstrating discriminative performance degradation. This degredation is primarily due to the infrastructure shift since $\Delta_{infra}>\Delta_{time}$.}
  \label{fig:performancegap} 
\end{figure}

\noindent
For simplicity we suppressed the display of $f$. \textbf{Figure \ref{fig:performancegap}} visualizes the breakdown of the AUROC ($\Delta_{\text{AUROC}}$) performance gap into $\Delta_{\text{AUROC}}^{time}$ and $\Delta_{\text{AUROC}}^{infra}$.

In terms of discriminative performance (AUROC), the differences in infrastructure pipelines between retrospective and prospective analyses had a larger impact on the performance gap compared to temporal shift. However, the converse is true for calibration (Brier score) gap, where the shift from \yrab{19}{20} to \yrab{20}{21} had a greater impact on calibration performance.

\begin{figure}[h]
\centering
    \subfigure[ 
    \label{fig:score_scatter}]{
        \includegraphics[width=2.75in]{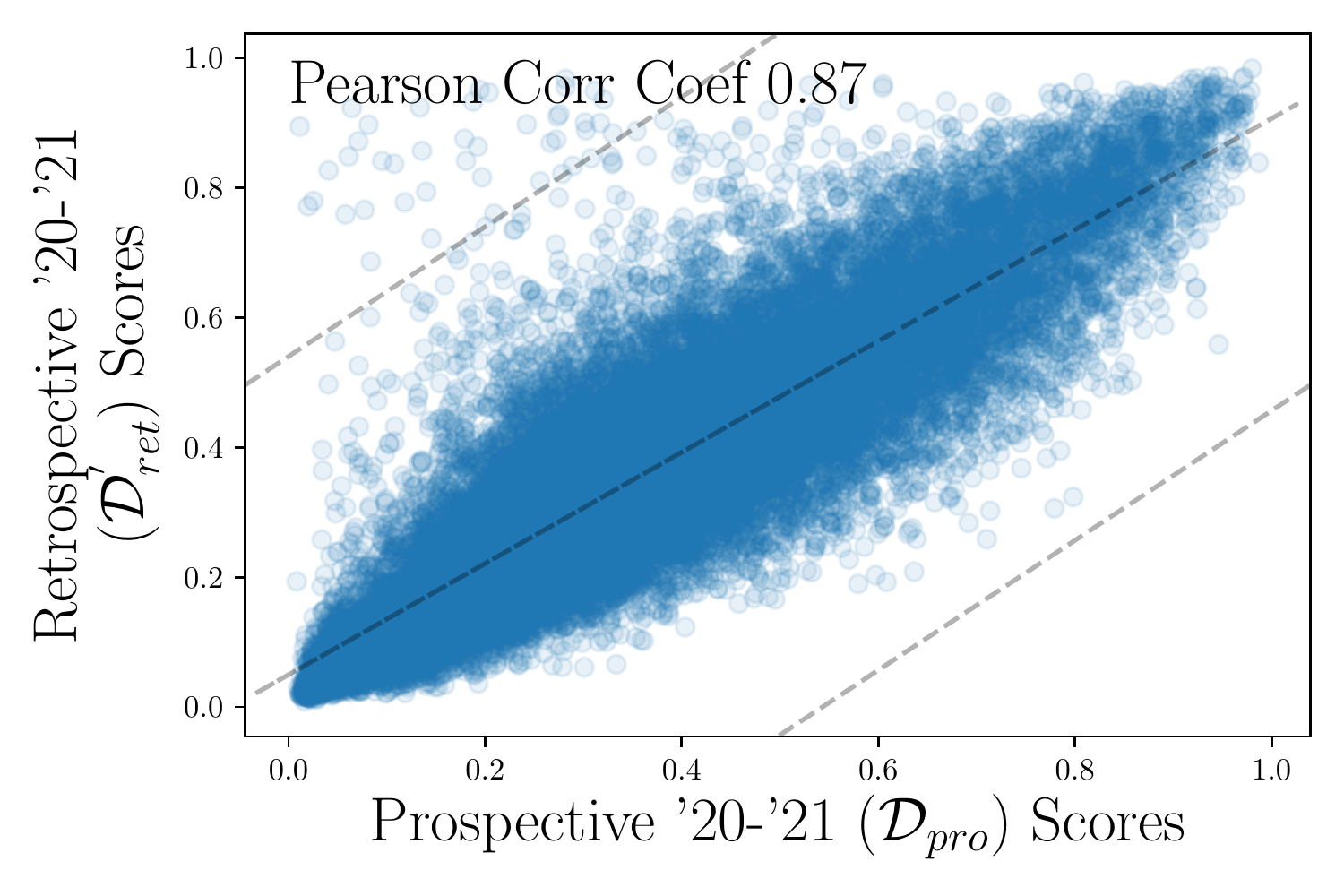}
    }
    \hspace{1em}
    \subfigure[ 
    \label{fig:discrepant_feature_histogram}]{
        \includegraphics[width=2.75in]{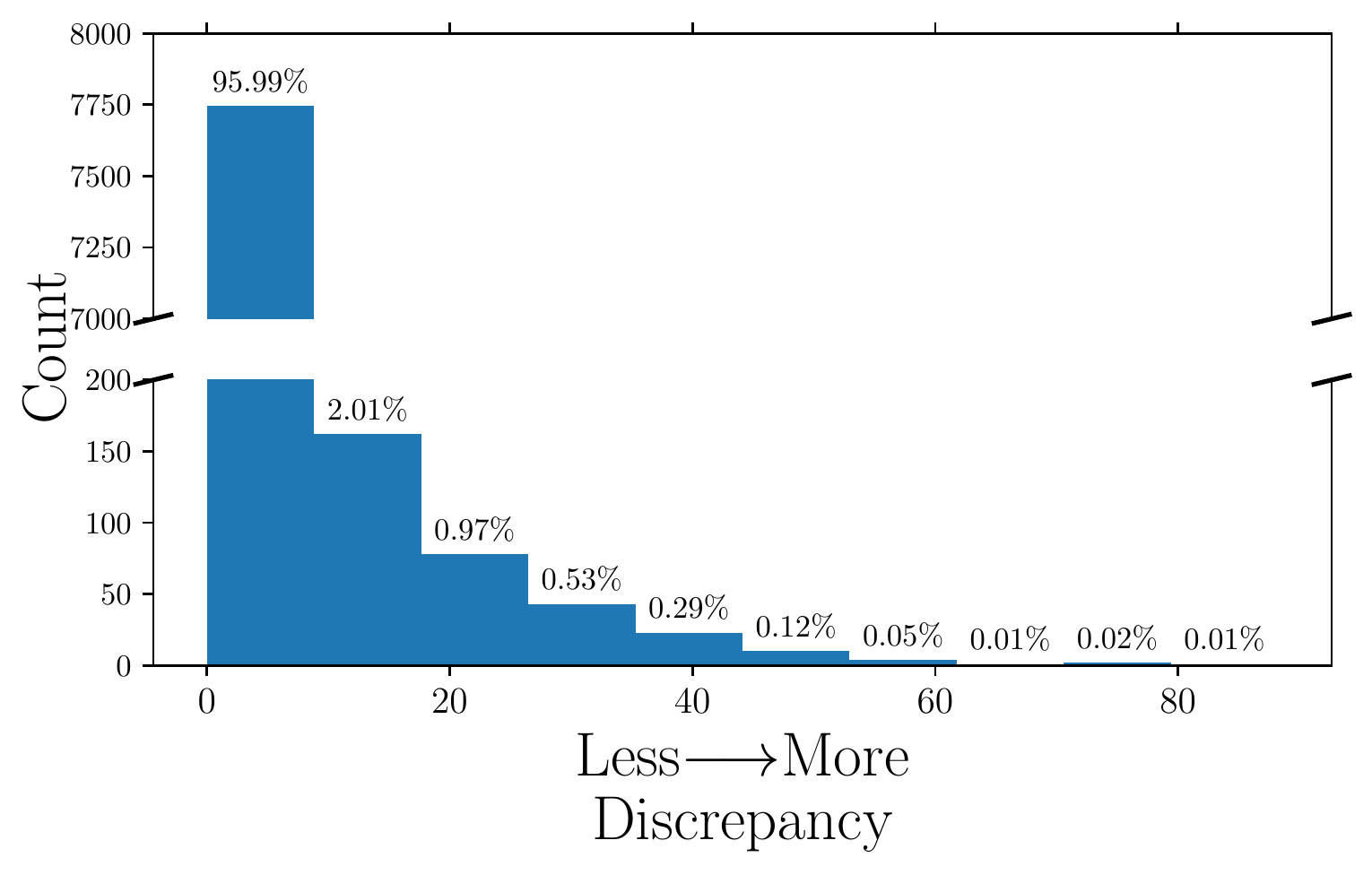}}

    \caption{Infrastructure Performance Gap Analysis. In (a) a scatter plot of risk scores generated by the \yrab{20}{21} prospective pipeline vs. \yrab{20}{21} retrospective pipeline is shown. We see that although highly correlated, the prospective and retrospective risk scores noticeably differ. In (b) we show the distribution of features based on how discrepant the features are (\textit{i.e.}, percent of instances where a feature is discrepant between retrospective and prospective \yrab{20}{21}). We see that although most features have low levels of discrepancy, there exists a subset of features whose values can vary greatly from the prospective to retrospective pipeline.}
    \label{fig:source_analyses}
\end{figure}

The infrastructure performance gaps indicate that the differences in the data extraction and processing pipeline led to a small (though not statistically significant) decrease in performance. When we compared the risk scores output by the model when applied to the retrospective versus prospective pipeline for every encounter in our \yrab{20}{21} cohort, we measured a correlation of 0.9 (\textbf{Figure \ref{fig:source_analyses}a}). 46 (0.2\%) encounters had extreme score differences (greater than $0.5$, denoted by the bounding dashed lines in the plot). 41 of these 46 encounters had a large number of days (more than 7 days for nearly all encounters) during which the prospective pipeline failed to run.

Comparing the input features, we found that 6,178 (77\%) of the 8,070 features had at least one instance (\ie encounter day) in which that feature differed across the two pipelines (\textbf{Figure \ref{fig:source_analyses}b}). However, only $1,612$ (20\%) of features differed in more than 1\% of instances. However, not all features are equally important. To measure the actual impact of the discrepancies on model performance, we must look at the feature swap analysis.

Applied to data from the first half of the prospective study period, the model achieved an AUROC performance of 0.769 on $\Dpro$. The AUROC after each feature group swap is displayed in \textbf{Table \ref{table:feature_swap}}. Hx: Medications, Idx: Medications, and Idx: In-Hospital Locations were the feature groups that had the largest positive swap difference in terms of AUROC, corresponding to improved model performance when given feature information from the retrospective pipeline. In the case of Hx: Medications, one would think these features would be consistent prospectively and retrospectively. However, we use both retrospective and prospective pipelines to calculate prospective values. To obtain 90 day patient histories we augment retrospective tables with prospective tables in order to fill the gap between when the data is logged in the EHR and when the data appears within RDW’s tables. In addition, to identify which previous admissions were inpatient admissions, we use patient class codes which are dynamic similar to laboratory results and medications. In addition to these more subtle changes, data that may be considered `static' (\eg where a patient lives or BMI) is liable to be change over the course of a patient encounter as information is collected and updated by the clinical care team. The full feature swap analysis is displayed in \textbf{Supplemental Table \ref{table:full_feature_swap}}.

\begin{table}[h]
    \centering
    \caption{Infrastructure Performance Gap Analysis - Feature Swap Performance. By swapping column values corresponding to feature groups between $\mathbf{X}_{pro}$ and $\mathbf{X}_{ret}'$ we were able to quantify the performance impact of differences in the infrastructure related to each feature group. Note, this analysis was conducted at an interim time-point of our study, as such only uses data from July 10th to December 21st for both $\Dpro$ and $\Dret'$. In addition to the feature group name, and the number of features in each feature group we display the AUROC on $\Dpro$ after the feature swap. Originally, we observed an AUROC of $0.769$ on $\Dpro$, the final column displays the difference between this value after the swap and the original $0.769$. We restrict this table to only positive differences, that is feature swaps that improve AUROC, all feature swap values are displayed in \textbf{Supplemental Table \ref{table:full_feature_swap} (Supplemental Material B)}. Hx: Medications, Idx: Medications, and Idx: In-Hospital Locations had the largest positive swap difference in terms of AUROC, corresponding to improved model performance when given feature information from the retrospective pipeline.}
    \label{table:feature_swap}
    \begin{tabular}{@{}lrr@{}}
    \toprule
    Feature Group & AUROC After Swap & Difference \\
    \midrule
    Hx: Medications                             & 0.787 &  0.018  \\
    Idx: Medications                            & 0.774 &  0.005  \\
    Idx: In-Hospital Locations                  & 0.772 &  0.003  \\
    Hx: Previous Encounters (Length of Stay)    & 0.770 &  0.001  \\
    Demographics: Body Mass Index               & 0.770 &  0.001  \\
    Demographics: County \& State               & 0.770 &  0.001  \\
    Idx: Colonization Pressure                  & 0.770 &  0.001  \\
    \bottomrule
    \end{tabular}
    \footnotesize{\\Descriptions of feature groups can be found in \textbf{Table \ref{table:feature_groups}}.}
\end{table}

\begin{table}[]
\centering
\caption{By feature group, lists the number of features that are significantly different between the \yrab{19}{20} study population and the \yrab{20}{21} study population. Significance was determined using a Z-test of the difference in proportions with a Bonferroni correction. One day was randomly sampled from each hospital encounter so that all feature instances are independent. Note, data that may be considered `obviously static', like the location a patient lives (\ie County \& State) may be updated over the course of an encounter, leading to discrepancies between prospective and retrospective data.}
\begin{tabular}{@{}lrr@{}}
\toprule
Feature Group & \begin{tabular}[c]{@{}l@{}}Number of Significantly\\ Different Features\end{tabular} & Total Number of Features \\ \midrule
Demographics                        & 0  & 124  \\
Hx: History of CDI                  & 0  & 2    \\
Hx: Diagnoses                       & 1  & 983  \\
Idx: Vital Sign Measurements        & 1  & 17   \\
Idx: Admission Details              & 5  & 22   \\
Hx: Previous Encounters             & 5  & 10   \\
Idx: Laboratory Results             & 6  & 508  \\
Idx: Colonization Pressure          & 7  & 10   \\
Hx: Medications                     & 23 & 2,731 \\
Idx: In-Hospital Locations          & 30 & 932  \\
Idx: Medications                    & 38 & 2,731 \\ \bottomrule
\end{tabular}
\label{table:temporal_shift}
\footnotesize{Descriptions of feature groups can be found in \textbf{Table \ref{table:feature_groups}}.}
\end{table}

Comparing the feature distributions between and $\Dret$ \yrab{19}{20} and $\Dret'$ \yrab{20}{21} we noted significant differences in 116 (1.44\%) of the features.  Features pertaining to medications, in-hospital locations, had the largest fraction of differences; however these categories also had a large number of overall features \textbf{Table \ref{table:temporal_shift}}. In terms of the fractions of differences within each category colonization pressure, patient history pertaining to number of previous encounters, and admission details had the greatest differences across time periods.

\section{Discussion \& Conclusion} 

In healthcare, risk stratification models are trained and validated using retrospective data that have undergone several transformations since being initially observed and recorded by the clinical care and operations teams. In contrast, during deployment models are applied prospectively to data collected in near real-time. Thus, relying on retrospective validation alone can result in an overestimation of how the model will perform in practice. In this paper, we sought to characterize the extent to which differences in how the data are extracted retrospectively versus prospectively contribute to a gap in model performance. We compared the performance of a patient risk stratification model when applied prospectively from July 2020-June 2021 to when it was applied retrospectively from July 2019-June 2020. Overall, the gap in performance was small. However, differences in infrastructure had a greater negative impact on discriminative performance compared to differences in patient populations and clinical workflows. 

 To date, much work has focused on addressing changes in model performance over time due to temporal shift  \citep{RN897, RN874, RN1004, RN1001}. In contrast, we focused on gaps due to differences in infrastructure. We relied on data extracted from a research data warehouse for model development and retrospective validation. Whereas, for near real-time prospective application of the model, we leveraged data extracted from a combination of custom web services and existing data sources. Prospectively, we had to shift to using the hospital census tables to identify our study cohort (\ie who was in the hospital) in real-time, in part because inpatient classification is dynamic and can shift over time. But even after accounting for differences in population, differences in how and when the data were sourced continued to contribute to a gap in performance.  Our analysis pointed to two sources of inconsistencies between the retrospective and prospective pipelines: i) inaccurate prospective infrastructure and ii) dynamic data entry. 

The first cause can be mitigated by revisiting the near real-time extraction, transformation, and load processes (\textit{e.g.}, rebuild prospective infrastructure to pull from different components of the EHR). For example, our analysis identified discrepancies in patient location codes between prospective and retrospective datasets. While the EHR passed the same location information to both pipelines, the two pipelines transformed and served this information in an inconsistent manner. Thus, we can rebuild the prospective infrastructure such that it uses the same processing code as the retrospective infrastructure.  The second cause is more difficult to address. The EHR itself is inherently dynamic as it serves many operational purposes \citep{RN887}. For example, the start and end dates for medications can change over time as the desired treatment plan changes and laboratory result names can change as initial results lead to further testing. In addition, specific aspects of features, such as laboratory result abnormality flags, can populate after the actual test results are filed (up to a day in our systems). To mitigate the impact of these differences on the performance gap, one can update the model to rely less on such dynamic elements of the EHR. For example, in our project, we substituted out medication orders for medication administrations. Although order time is available earlier than administration, orders are frequently cancelled or updated after they are initially ordered by a physician. In contrast, medication administration information is more stable across time. Our findings underscore the need to build pipelines that are representative of the data available at inference time. The closer the retrospective representation of data are to data observed prospectively, the smaller the potential performance gap.


Beyond differences in infrastructure, it is reassuring that changes in patient populations and workflows between time periods (\ie temporal shift) did not increase the gap in discriminative performance. On a month-by-month basis the only significant differences in performance were during the months of March and May, otherwise the model performed as well, if not better, prospectively. Interestingly, predicting which patients were most likely to acquire CDI during the current hospital visit was significantly easier in March 2020, compared to March 2021. This discrepancy is likely due to significant operational changes at University of Michigan Health due to the onset of the COVID-19 pandemic. Comparing the expected feature vectors in \yrab{19}{20} vs. \yrab{20}{21}, we noted significant differences in locations and admission types, changes likely attributed to the COVID-19 pandemic. For example, new patient care units were created for patients hospitalized for COVID-19 \citep{RN994} and patient volume to existing units and services decreased significantly \citep{RN997, RN998, RN999}.  Additionally, colonization pressure depends on locations, as such we would expect this to change with the distribution of patients in locations changing. Many of the other changes in feature groups may also be explained by this drastic change in patient population. While these changes actually made the problem easier during prospective validation (\textbf{Supplemental Material C}), in-line with previous work, the calibration performance of the model was negatively impacted by the temporal shift \citep{RN841, RN1001, RN874}. 

This study is not without limitations. Aside from the  limitations associated with studying a single model at a single center, there is another nuanced limitation that pertains to timing of data. The age (\ie time between data collection and use for this analysis) of the retrospective data varied in our analysis. Some validation data had been present in RDW for over two years, while other data were populated far more recently. Data collected in large retrospective databases are always subject to change, but the likelihood of changes decreases over time as updates from clinical care and billing workflows settle. As we use data closer to the present (July 2021), it is possible that the data may continue to change. Thus, if we were to revisit the analysis in the future, the infrastructure gap could further increase. However, most updates to the retrospective data occur within 30 days of discharge and thus we expect the impact on our results to be limited.

The performance gap is due, in part, to the fact that we are trying to capture a moving target with a single snapshot. Existing EHR and associated database systems are primarily designed to support care operations. Therefore, they lack features to help with development and deployment of real-time predictive models. EHR vendors are working to develop tools for the efficient and effective deployment of ML models \citep{RN864}. However, to the extent that we continue to develop models using data extracted from databases that are several steps removed from clinical operations, issues are likely to remain. While overwriting records and values may be of little consequence for care operations, it makes retrospective training fraught with workflow issues. Mechanisms are needed to keep track of what the database looked like at every moment in time - \`a la Netflix's time machine \citep{RN560}. However, in lieu of such solutions, thorough prospective validation and analysis can help bridge the gap, providing a more accurate evaluation of production model behavior and elucidating areas for improvement. 

\acks{This work was supported by the National Science Foundation (NSF award no. IIS-1553146) the National Institute of Allergy and Infectious Diseases of the National Institutes of Health (grant no. U01AI124255) and the Agency of Health Research and Quality (grant no. R01HS027431-02) and Precision Health at the University of Michigan (U-M).
The authors would like to thank the U-M Data Office, including Erin Kaleba and Robinson Seda, Precision Health, including Sachin Kheterpal, Hae Mi Choe, Jessica Virzi, and Susan Hollar, and the entire U-M Research Data Warehouse team, including Jeremy Jared and Peter Bow, among others.}

\bibliography{bibliography/20210716}


\newpage
\section*{Supplemental Material}
\subsection*{Supplemental Material A - Feature Groups}

Feature groups are grouping of features that have a shared meaning or source. We utilize them to pin-point sources of discrepancies between the prospective and retrospective pipelines. Feature groups may be composed of other feature groups, \textbf{Table \ref{table:feature_groups}} displays this hierarchical aspect of feature groups. For example, feature group ``Idx: Admission Details'' contains the feature groups ``Idx: Admission Type'' and ``Idx: Insurance Type''.

\begin{center}
\begin{longtable}{llr}
\caption{Feature Groups and their descriptions. Feature groups are hierarchical, with two major categories \textbf{Demographics} and \textbf{Clinical Characteristics}. Demographics are generally static patient-level attributes. Clinical characteristics are dependent on time and may change over the course of an encounter. Clinical characteristics may also be broken into two major categories, based on which encounters the information is tied to: \textit{historical encounters} or \textit{index encounters}. Historical encounter information may be denoted in the main text with a ``Hx" prefix and represents information collected in the encounters leading up to the current encounter. This history look-back is limited to 90 days. The index encounter information pertains to the current encounter and may be denoted with the ``Id'' prefix.  Descriptions of each of these feature groups are provided along with the number of features included in this feature group. Various levels of feature group hierarchical structure are employed depending on the analysis.}
\label{table:feature_groups}
\\
\hline
Feature Group & &Number of Features \\ 
\hline
\endfirsthead
\multicolumn{3}{c}%
{\tablename\ \thetable\ -- \textit{Continued from previous page}} \\
\hline
Feature Group & &Number of Features \\ 
\hline
\endhead
\hline \multicolumn{3}{c}{\textit{Continued on next page}} \\
\endfoot
\hline
\endlastfoot
\textbf{Demographics} & & 124 \\
\quad Age & & 5\\
\quad Gender & & 2\\
\quad Race & & 8\\
\quad Marital Status & & 2\\
\quad County \& State & & 102\\
\quad Body Mass Index & & 5\\

\textbf{Clinical Characteristics} & & \\

\quad\textit{Historical Encounters (Hx)} & & \\
\quad\quad History of CDI & & 2\\
\quad\quad Previous Encounters (stats) & & 10\\
\quad\quad\quad Number of Previous Encounters & & 3\\
\quad\quad\quad Length of Stay & & 7\\
\quad\quad Diagnoses (Diagnosis-Related Group/ICD9/ICD10) & & 983\\
\quad\quad Medications & & 2,731\\
\quad\quad\quad Medication & & 1,886\\
\quad\quad\quad Ingredient & & 620\\
\quad\quad\quad Class & & 225\\

\quad \textit{Index Encounter (Idx)} & & \\
\quad\quad Admission Details & & 22\\
\quad\quad\quad Admission Type & & 3\\
\quad\quad\quad Patient Type & & 12\\
\quad\quad\quad Insurance Type & & 6\\
\quad\quad\quad Emergency Visit & & 1\\
\quad\quad In-Hospital Locations & & 932\\
\quad\quad Vital Sign Measurements & & 17\\
\quad\quad Laboratory Results & & 508\\ 
\quad\quad Medications & & 2,731\\
\quad\quad\quad Medication & & 1,879\\
\quad\quad\quad Ingredient & & 629\\
\quad\quad\quad Class & & 223\\
\quad\quad Colonization Pressure & & 10\\ 
\quad\quad\quad Unit-based & & 5\\
\quad\quad\quad Hospital-wide & & 5\\
\hline
\end{longtable}
\end{center}

\newpage
\subsection*{Supplemental Material B - Infrastructure Performance Gap Analysis}

\begin{table}[h]
    \centering
    \caption{Infrastructure Performance Gap Analysis - Full Feature Swap Performance. By swapping column values corresponding to feature groups between swap analysis between $\mathbf{X}_{pro}$ and $\mathbf{X}_{ret}'$ we were able to quantify the performance impact of differences in the infrastructure related to each feature group. Note, this analysis was conducted at an interim time-point of our study, as such only uses data from July 10th to December 21st for both $\Dpro$ and $\Dret'$. In addition to the feature group name, and the number of features in each feature group we display the AUROC on $\Dpro$ after the feature swap. Originally, we observed an AUROC of $0.769$ on $\Dpro$, the final column displays the difference between this value after the swap and the $0.769$. Hx: Medications, Idx: Medications, and In-Hospital Locations were the feature groups that had the largest positive swap difference in terms of AUROC, corresponding to improved model performance when given feature information from the retrospective pipeline.}
    \label{table:full_feature_swap}
    \begin{tabular}{@{}lrr@{}}
    \toprule
    Feature Category & AUROC After Swap & Difference \\
    \midrule
    Hx: Medications                                         & 0.787 &  0.018  \\
    Idx: Medications                                        & 0.774 &  0.005  \\
    Idx: In-Hospital Locations                              & 0.772 &  0.003  \\
    Hx: Previous Encounters (Length of Stay)                           & 0.770 &  0.001  \\
    Demographics (Body Mass Index)                                      & 0.770 &  0.001  \\
    Demographics (County \& State)                          & 0.770 &  0.001  \\
    Idx: Colonization Pressure                              & 0.770 &  0.001  \\
    Demographics (Race)                                     & 0.769 &  0.000  \\
    Idx: Admission Details (Emergency Visit)                & 0.769 &  0.000  \\
    Demographics (Gender)                                   & 0.769 &  0.000  \\
    Hx: History of CDI                                      & 0.769 &  0.000  \\
    Idx: Admission Details (Patient Type)                   & 0.769 &  0.000  \\
    Demographics (Age)                                      & 0.769 &  0.000  \\
    Demographics (Marital Status)                           & 0.769 &  0.000  \\
    Idx: Admission Details (Admission Type)                 & 0.769 &  0.000  \\
    Idx: Admission Details (Insurance Type)                 & 0.769 &  0.000  \\
    Hx: Previous Encounters (Number of Previous Encounters) & 0.769 &  0.000  \\
    Idx: Laboratory Results                                 & 0.768 & -0.001  \\
    Hx: Diagnoses                                           & 0.767 & -0.002  \\
    Idx: Vitals                                             & 0.766 & -0.003  \\
    \bottomrule
    \end{tabular}
    \footnotesize{Descriptions of feature groups can be found in \textbf{Table \ref{table:feature_groups}}.}
\end{table}

\newpage
\subsection*{Supplemental Material C - Model Performance Pre vs During COVID-19}

\begin{figure}[h]
  \centering 
  \includegraphics[width=4.0in]{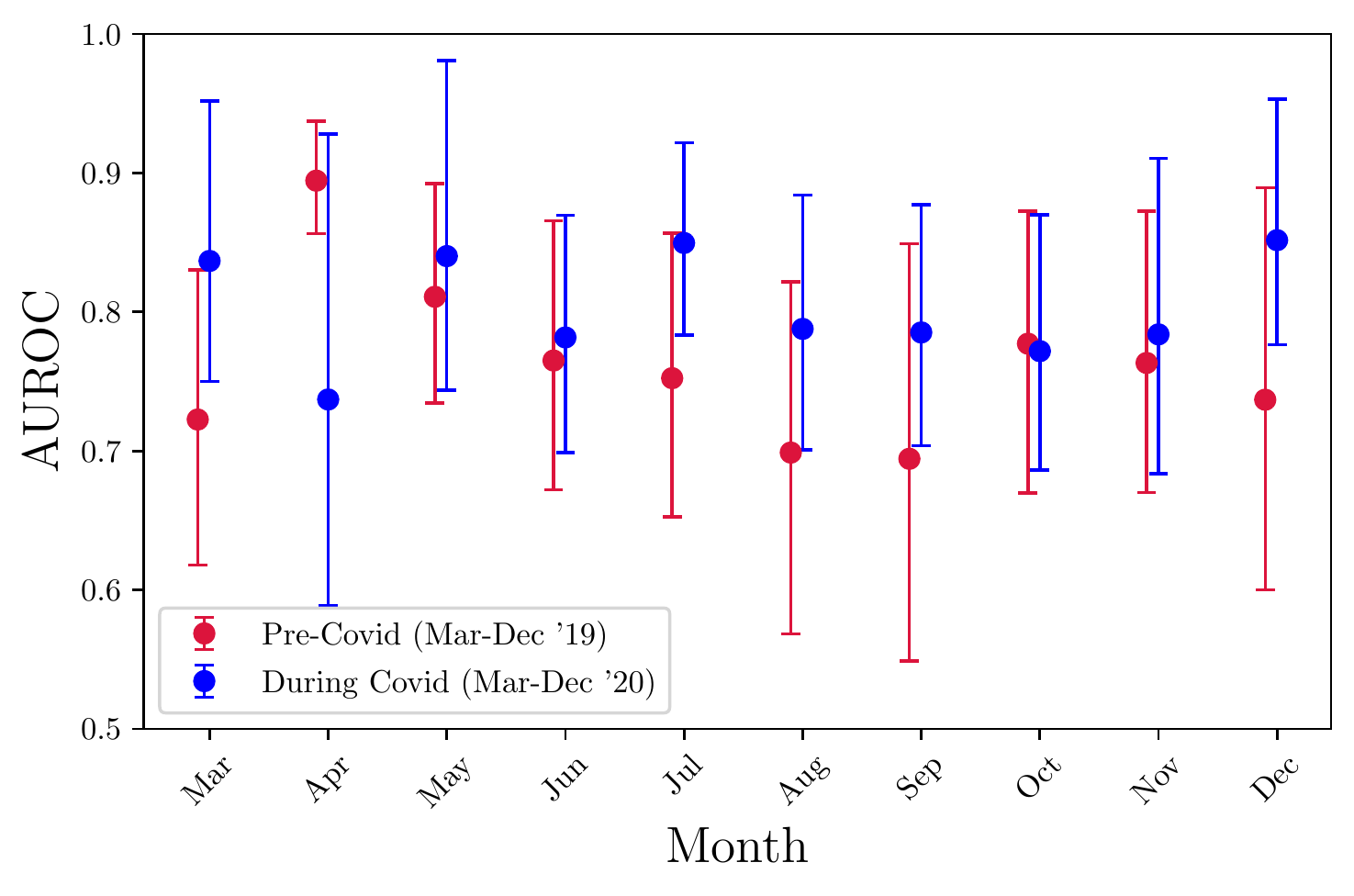} 
  \caption{Monthly performance of model in 2019 vs 2020 to show trends in performance before and during COVID-19. We see that performance is slightly higher during COVID-19.}
  \label{fig:covid} 
\end{figure}

In order to measure the impact of COVID-19 on model performance, we look at monthly AUROC performance before and during COVID-19 in \textbf{Figure \ref{fig:covid}}. We notice that performance pre-Covid is generally lower than during Covid with the exception of April which has long error bars due to small numbers of cases. We hypothesize the improved performance during Covid may be due to a simplification of the task. Our task is to predict hospital associated, hospital-associated CDI. However, our method for distinguishing between hospital associated, hospital-associated vs community associated/recurrent, hospital-associated is dictated by guidelines which may or may not always reflect ground truth. We hypothesize that the increased contact precautions led to relatively fewer hospital associated cases and relatively more community associated/recurrent cases. The latter is easier to identify because it only requires identifying susceptibility versus susceptibility and exposure. 

\newpage
\subsection*{Supplemental Material D - \yrab{18}{19} Retrospective Validation}

\begin{figure}[h]
\centering
    \subfigure[ROC Curve with 95\% Confidence Interval. 
    \label{fig:validation_rocs_1819}]{
        \includegraphics[width=2.75in]{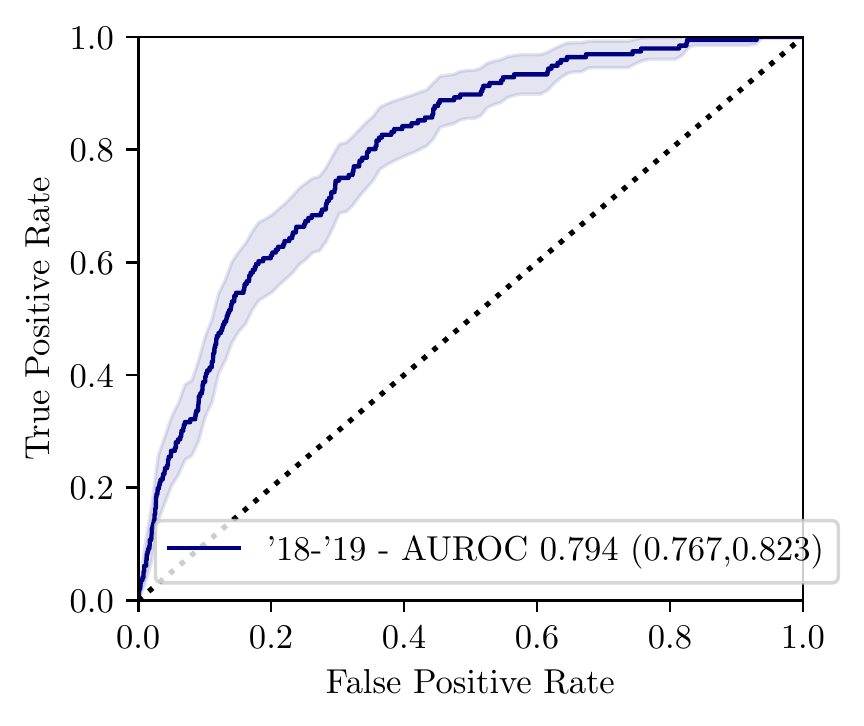}}
    \hspace{1em}{}
    \subfigure[Confusion Matrix and Performance Measures.  
    \label{fig:validation_confusion_matrices_1819}]{
        \includegraphics[width=1.25in]{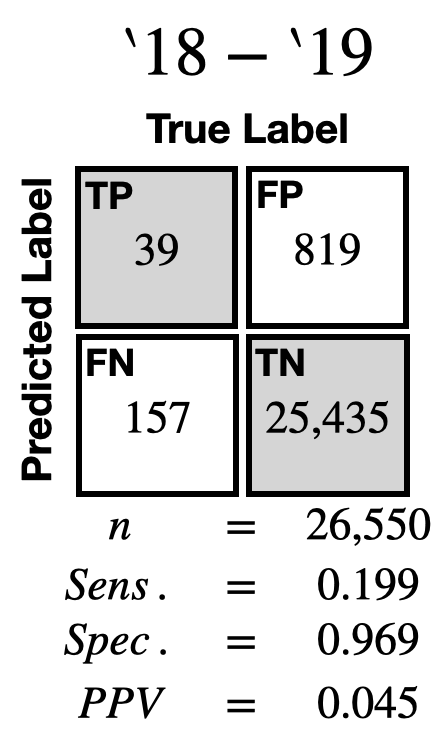}
    }
    \caption{Risk prediction model performance on the retrospective \yrab{18}{19} validation dataset.}
    \label{fig:validation_performance_1819}
\end{figure}

\begin{figure}[h]
  \centering 
  \includegraphics[width=4.0in]{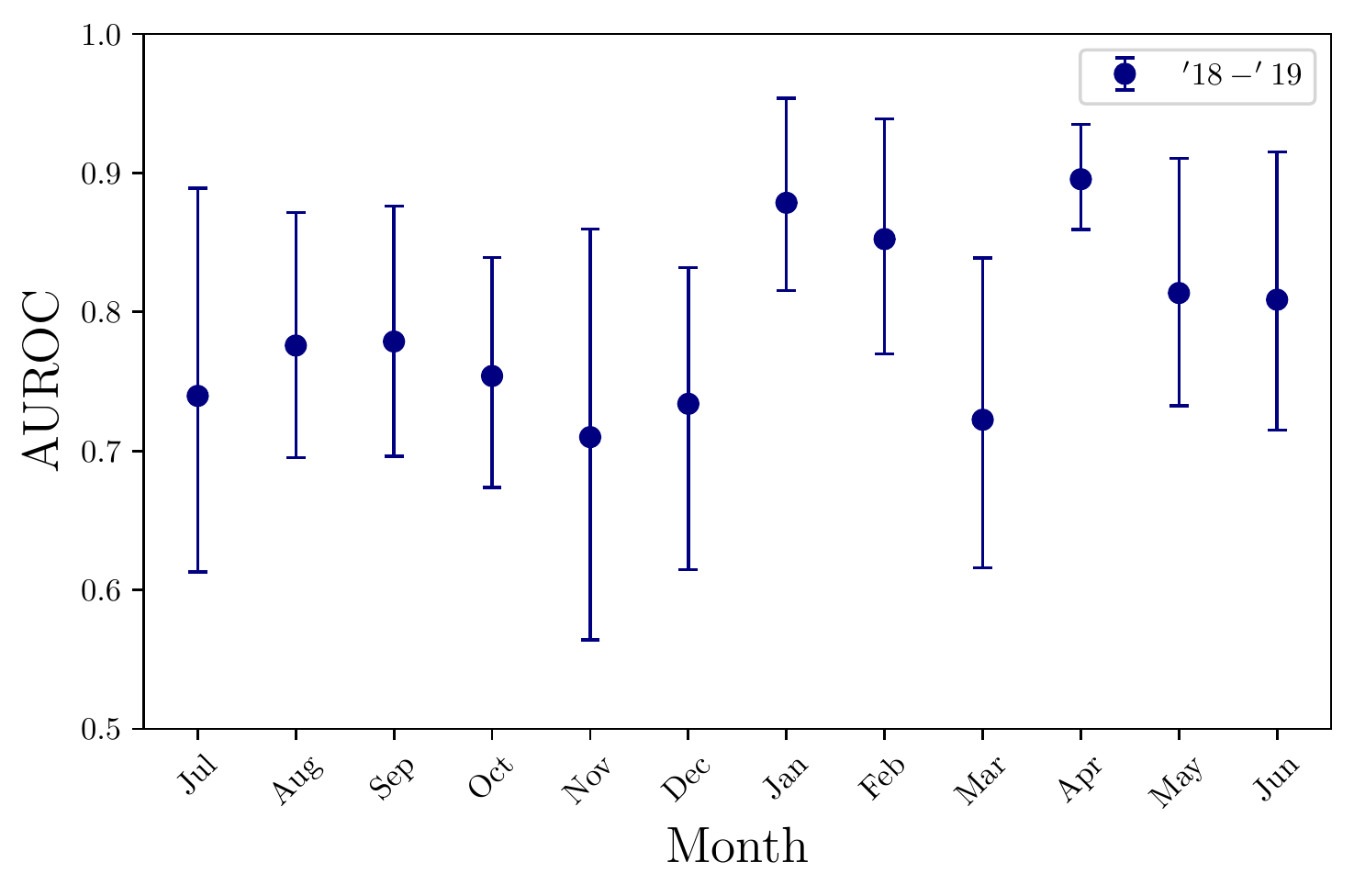} 
  \caption{Monthly AUROC Performance. AUROC for \yrab{20}{21} prospective dataset and the \yrab{19}{20} retrospective dataset broken down by month and bootstrap sampled $1,000$ times to generate 95\% confidence intervals. We see that performance fluctuates month-by-month with higher performance in January, February and April. There appear to be some monthly trends in performance across years. Similar to \yrab{19}{20} we see lower, less variable scores, in the later months of the year with higher and more variable scores in the earlier months of the year.}
  \label{fig:monthlyauroc_1819} 
\end{figure}

All validation datasets include encounters from July 10th to June 30th of the following year. After applying inclusion criteria, the \yrab{18}{19} retrospective validation set consisted of 26,450 hospital encounters, population characteristics are detailed in \textbf{Table \ref{table:pop_characteristics_1819}}. It should be noted that the \yrab{18}{19} time period overlaps with the model validation period in 2018. This means that feature distributions from 2018 were used to help inform the decision to discard rare features. Applied to the retrospective validation data from \yrab{18}{19} the risk prediction model achieved AUROCs of $0.794$ (\ci$0.767$, $0.823$) (\textbf{Figure \ref{fig:validation_performance_1819}a}). Selecting a decision threshold based on the 95th percentile of risk from the training set and applying on \yrab{18}{19} led to positive predictive values of $0.045$, $0.036$, and $0.027$ respectively  (\textbf{Figure \ref{fig:validation_performance_1819}b}). Monthly performance for \yrab{18}{19} is displayed in \textbf{Figure \ref{fig:monthlyauroc_1819}}.

\begin{table}[h]
    \centering
    \caption{\yrab{18}{19} Cohort Characteristics.}
        \label{table:pop_characteristics_1819}
    \begin{tabular}{@{}ll@{}}
    \toprule
                    & \yrab{18}{19} \\ 
                    & n=26,450 \\
    \midrule
    Median Age (IQR)                        & 59 (41, 70)  \\
    Female (\%)                             & 51\%         \\
    Median Length of Stay (IQR)                        & 5 (4, 9)     \\
    History of CDI in the past year (\%)    & 1.7\%        \\
    Incidence Rate of CDI (\%)              & 0.7\%        \\
    \bottomrule
    \end{tabular}
\end{table}

\end{document}